\documentclass[11pt]{article}

\def\be{\begin{eqnarray}}
\def\ee{\end{eqnarray}}
\def\bq{\begin{equation}}
\def\eq{\end{equation}}
\def\ben{\begin{enumerate}}\def\een{\end{enumerate}}

\def\roughly#1{\mathrel{\raise.3ex\hbox{$#1$\kern-.75em%
\lower1ex\hbox{$\sim$}}}}

\usepackage{epsfig,cite}

\begin{document}
\begin{titlepage}

\hfill FTUV-15-0225

\hfill IFIC/15-29

 \vspace{1.5cm}
\begin{center}
\ \\
{\bf\LARGE Glueball-Meson Mixing}

\vspace{0.7cm} {\bf\large Vicente Vento} \vskip 0.7cm

\end{center}

\noindent  {Departamento de F\'{\i}sica Te\'orica and Instituto de
F\'{\i}sica Corpuscular, Universidad de Valencia - \\
Consejo Superior de Investigaciones Cient\'{\i}ficas, 46100 Burjassot (Val\`encia), Spain, \\{\small Email: vicente.vento@uv.es}.}

\vskip 1cm \centerline{\bf Abstract} \vskip 0.3cm
Calculations in unquenched QCD for the scalar glueball spectrum have confirmed previous results of Gluodynamics finding a glueball at $ \sim 1750 $ MeV. I analyze the implications of this discovery
from the point of view of  glueball - meson mixing at the light of the experimental scalar sprectrum.\vspace{2cm}

\noindent Pacs: 12.38.-t, 12.38.Aw,14.70.-e, 14.80.-j


\end{titlepage}

\section{Introduction}

\indent\indent

Glueballs have not been an easy subject to study due to the lack
of phenomenological support and therefore much debate has been
associated with their properties \cite{Mathieu:2008me}.  I center the
 discussion here at the consequences of the spectrum obtained by  lattice QCD where the results seem to be converging.
In quenched QCD  the masses of the scalar glueballs appear large $ \ge 1700$ MeV
\cite{Lee:1999kv,Vaccarino:1999ku, Morningstar:1999rf, Chen:2005mg}, a result which has been confirmed 
 by unquenched calculations \cite{Gregory:2012hu}. . 

Several glueball-meson mixing scenarios have been discussed in the literature using either lattice calculations or phenomenology \cite{Lee:1999kv,Vaccarino:1999ku,Amsler:1995td,Barberis:1996iq,Cheng:2009dg}.  I implement a combination of lattice results and phenomenology to study the implications of recent lattice results in the possible mixing scenarios in the scalar sector.

\section{The scalar spectrum}

Lattice QCD provides us with a value for the mass of the $0^{++}$ glueball states as shown in Table \ref{gmass}.

\begin{table}[ht]
\begin{center}
  \begin{tabular}[h]{|c|c|c|c|}
\hline
$J^{PC}$  & \multicolumn{3}{|c|}{Mass MeV} \\ 
\cline{2-4}
 & Unquenched & \multicolumn{2}{|c|}{Quenched } \\
\cline{2-4}
 & Gl&  Mp & Ky \\  \hline
%%%
$0^{++}$ & 1795(60)  & 1730(50)(80)   & 1710(50)(80) \\ \hline 
$0^{++}$ & 3760(240) & 2670(180)(130) &   \\ 
\hline
  \end{tabular}
\end{center}
\caption{Glueball masses with $J^{PC}$ assignments.
The column Gl reports the results of the unquenched  QCD calculation by Gregory et al.~\cite{Gregory:2012hu}, the columns Mp and Ky show the data from 
Morningstar and Peardon~\cite{Morningstar:1999rf}
and  Chen et al.~\cite{Chen:2005mg} respectively.
}
\label{gmass}
\end{table}

The three calculations give a similar mass for their lowest state for which I take the mean $1743 \pm 42 $ MeV in my analysis.

In Table \ref{pdg}, I show the experimental scalar spectrum, namely that of the particles labelled $f_0$.

\begin{table}[ht]
\begin{center}
  \begin{tabular}[h]{|c|c|c|c|c|}
\hline
$J^{PC}$  & Name &  {Mass MeV} & {Width MeV} & Comment\\ \hline
%%%
$0^{++}$ & $f_0(500)$ & 400-550 &  400-700&\\ \hline
$0^{++}$ & $f_0(980)$ & 990$ \pm $20 &  40-100&\\ \hline
$0^{++}$ & $f_0(1370)$& 1200-1500 &  200-500&\\ \hline
$0^{++}$ & $f_0(1500)$& 1505 $\pm$ 6  &  109 $\pm$ 7& \\ \hline
$0^{++}$ & $f_0(1710)$& 1720 $\pm$ 6 &  135 $\pm$ 8&\\ \hline
$0^{++}$ & $f_0(2020)$&  1992 $\pm$ 16 & 442 $\pm$ 60 & needs confirmation \\ \hline
$0^{++}$ & $f_0(2100)$&  2013 $\pm $ 8 & 209 $\pm$ 19 & needs confirmation\\ \hline
$0^{++}$ & $f_0(2200)$&  2189 $\pm$ 13 & 238 $\pm$50 & needs confirmation \\ \hline
  \end{tabular}
\end{center}
\caption{Scalar particles appearing in the PDG's particle listings \cite{Agashe:2014kda}.
}
\label{pdg}
\end{table}

The three heaviest states have not been confirmed and some authors also question the existence of the  $f_0 (1370)$.

An observation at the light of the experimental spectrum  is that the excited glueballs obtained in QCD calculations are very high in mass and therefore I do not expect them to mix with the mesons. Thus the lattice calculations and the observed scalar spectrum lead to a scenario of one glueball amidst several scalar mesons in the range between $1-2$ GeV.

\section{Glueball-Meson Mixing}

The most naive way of implementing a mixing scenario is by assuming a hamiltonian which is not diagonal in the unmixed states. Thus there are two ingredients to this hamiltonian, the unmixed masses and the off-diagonal mixing parameters.  

The calculations thus far have led to different realizations of mixing. Some obtain that the  $f_0 (1710)$ is mostly glueball with small admixtures of $q\bar{q}$ states \cite{Lee:1999kv,Vaccarino:1999ku,Cheng:2009dg}. Others claim that the $f_0(1500)$ is mostly glueball, with small admixtures of  $q\bar{q}$ states \cite{Amsler:1995td,Barberis:1996iq}.

%%%%%%%%%%%%%%%%%%%%%%%%%%%%%%%%%%%%%%%%%%%%%%%%%%%%%%%%%
\vskip 0.3cm
\begin{figure}[htb]
\begin{center}
\epsfig{file=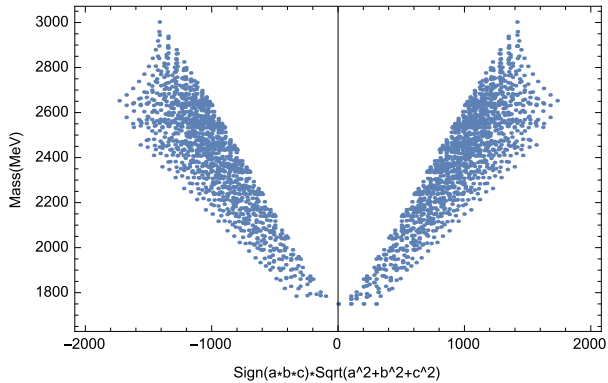,width=8.0cm,angle=0}
\caption{The largest eigenvalue of the studied analog of a three state mixing matrix is shown as a function of the relevant parameters in Cardano's formulas. The values used for the diagonal elements were $1750, 1500, 1300$  close to the physical $f_0$ masses, and the mixing parameters $b,c$ range from $0 \to1000$, and $a$ from $ -1000\to1000$ in discrete steps.}
\label{cardano}
\end{center}
\end{figure}

%%%%%%%%%%%%%%%%%%%%%%%%%%%%%%%%%%%%%%%%%%%%%%%%%%%%%%%%%

I  analyze next which scenarios could arise in a full QCD calculation.  The idea is to construct a hamiltonian which is not diagonal in the unmixed lattice states and diagonalize it to obtain the physical states. The decisive input is the glueball mass as given by lattice QCD. I assume that the rest of the particles in the unmixed Fock space are scalar mesons. Since the scalar meson spectrum of lattice QCD is not well known I will be guided by phenomenology.

For the off-diagonal matrix elements I shall use perturbation theory and  $1/N_c$ arguments. With these assumptions the matrix elements are of the form $ \sim |<\Psi_n|H_{non-diagonal}|\Psi_m>|^2/(E_n-E_m)$, therefore they depend on the inverse of the decay coupling $f_{meson}^2 \sim N_c$ , $ f_{glueball}^2 \sim N_c^2$, and on the inverse of the energy difference of the two levels. For almost degenerate unmixed states, the mixing parameters might be larger than for states of quite different masses. Thus the hamiltonian matrix might look  for  a three state mixing as,

 \begin{equation}  \left( 
 \begin{array}{c c c }
 %%%
  m_{s1} & a & b \\
  a&m_{s2}  & c   \\
  b & c & m_g \\
  \end{array}   \right) ,
  \label{matrix}
  \end{equation}\\[0.3cm]
where,  $m_{s1}$ and $ m_{s2}$ represent the unmixed scalar meson masses,  and $m_{g}$  the unmixed glueball mass. The diagonal matrix elements I take motivated by phenomenology and their values incorporate all higher order corrections, $\sim 1 + O(N_c^{-2})+\ldots$ for the glueballs, and $\sim 1 + O(N_c^{-1})+\ldots$ for the mesons. The mixing matrix elements are $\sim O(N_c^{-2}) $ for meson-glueball states since $\Delta E \sim 1/N_c$, but they are also  $ \sim O(N_c^{-2} )$ for the meson-meson states since $\Delta E \sim O(1)$ for them.  Thus $a,b$ and $c$  will be of the same order of magnitude, i.e.  $\le 250 $ MeV. 

It is trivial to show that in a two by two symmetric mass mixing matrix the the unmixed mass of the heavy state will increase after non trivial diagonalisation. For  a three state mass mixing matrix it is not so trivial to show that the heaviest state mass also increases after non trivial diagonalisation. This can be done using Cardano's formulas, as shown in Fig. \ref{cardano}. Cardano's formulas depend only of one sign, $sign(abc)$, therefore one reaches all the desired values by only letting one of the parameters become negative. I have used as diagonal elements numbers close to the actual $f_0$ masses and the off diagonal elements have taken values far beyond my expectation for the physical mixing parameter values. 

%%%%%%%%%%%%%%%%%%%%%%%%%%%%%%%%%%%%%%%%%%%%%%%%%%%%%%%%%
\vskip 0.3cm
\begin{figure}[htb]
\begin{center}
\epsfig{file=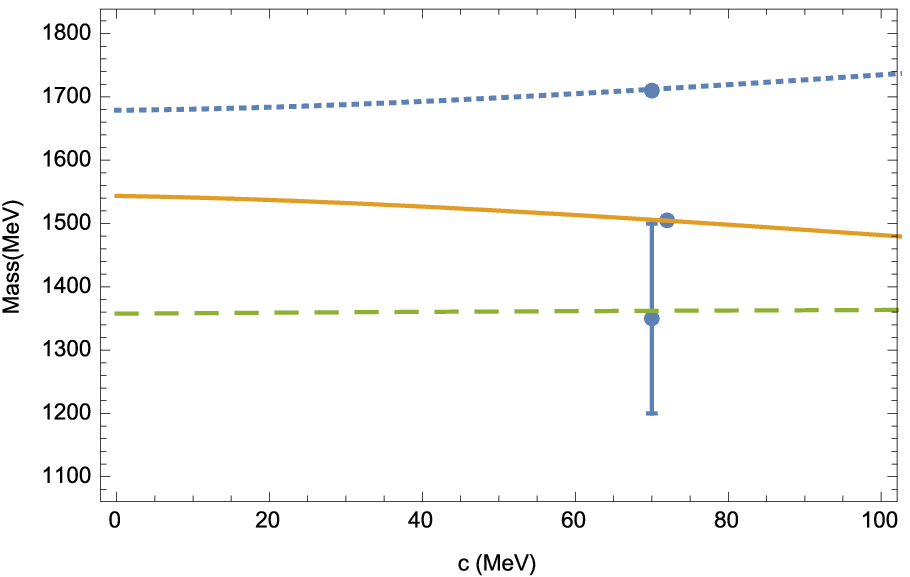,width=6.5cm,angle=0} \hspace{0.8cm}
\epsfig{file=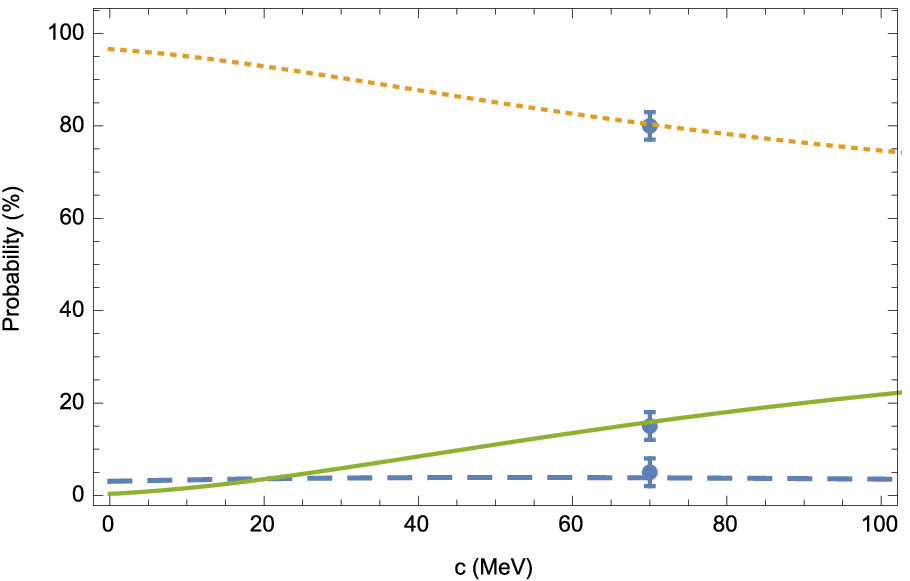,width=6.5cm,angle=0} 
\caption{Left: Eigenvalues of the mixing matrix for the glueball and two meson states.  The masses and their errors are from PDG \cite{Agashe:2014kda}. Right: The probability distribution  for the $f_0(1710)$ of the unmixed states  as a function of $c$, having fixed  $a=b=50$ MeV. The dotted line represents the glueball with unmixed mass ($1670$ MeV), the full line the heavy meson with unmixed mass ($1530$ MeV), the dashed line the light meson with unmixed mass ($\sim 1380$ MeV). The error bars in the probability curves are just to guide the eye.}
\label{threeMP}
\end{center}
\end{figure}
%%%%%%%%%%%%%%%%%%%%%%%%%%%%%%%%%%%%%%%%%%%%%%%%%%%%%%%%%

The QCD lattice mass value for the scalar glueball is very high and the error relatively small if the three values shown in Table \ref{gmass} are averaged with errors in quadrature  $m_g=1743 \pm 42$ MeV. One can still argue, given the errors in the lattice calculation, that the glueball mass could be below, but certainly not much below, the experimental $f_0 (1710)$. Therefore, I  foresee two scenarios: i) the unmixed QCD mass value is below the $f_0 (1710)$ but close to it, and ii) the unmixed glueball mass value is  above  the $f_0(1710)$ in agreement with the lattice QCD average value. In the latter case  I will analyze two cases depending on the unmixed spectrum, one will be associated with weak mixing, while the other with strong mixing. 

%%%%%%%%%%%%%%%%%%%%%%%%%%%%%%%%%%%%%%%%%%%%%%%%%%%%%%%%%
\vskip 0.3cm
\begin{figure}[htb]
\begin{center}
\epsfig{file=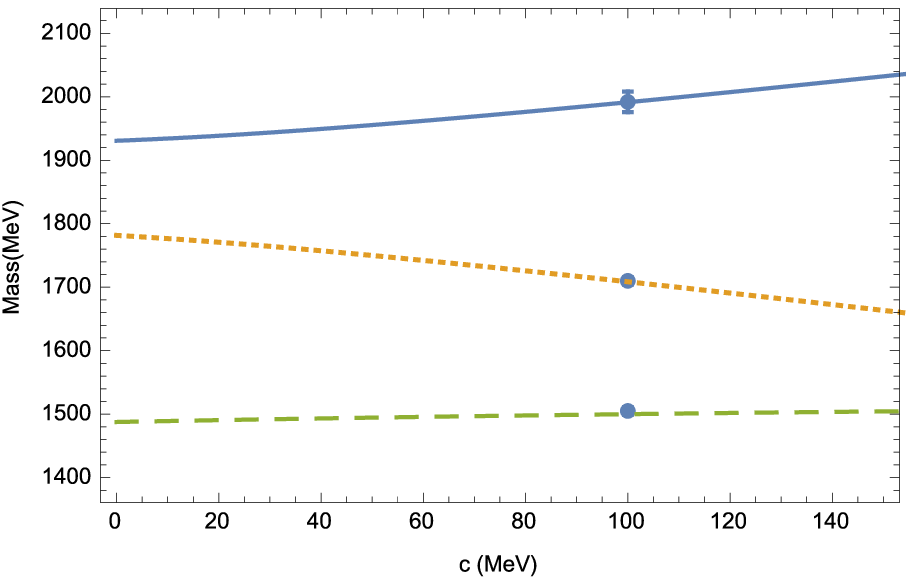,width=6.5cm,angle=0} \hspace{0.8cm}
\epsfig{file=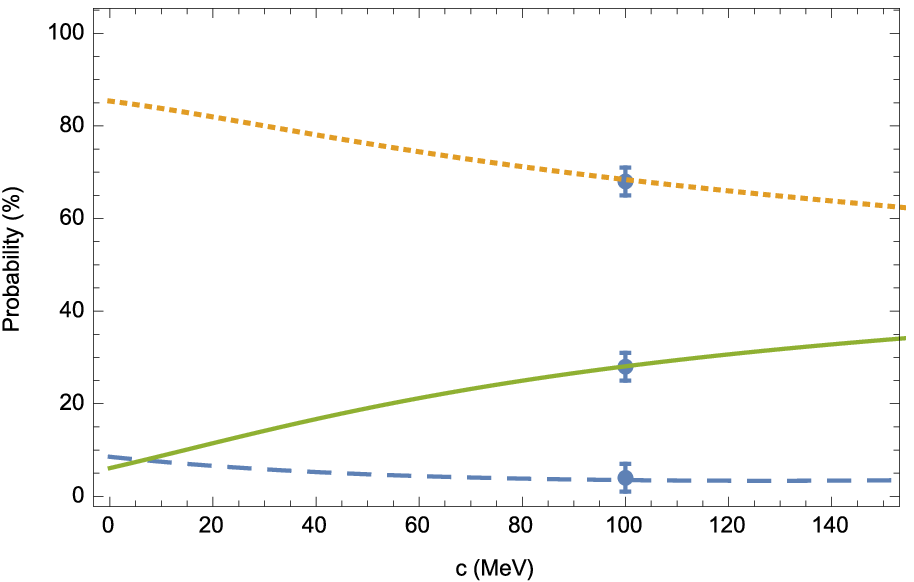,width=6.5cm,angle=0} 
\caption{Left: Eigenvalues of the mixing matrix for the glueball and two meson states.  The masses and their errors are from PDG \cite{Agashe:2014kda}. Right: The probability distribution for the $f_0(1710)$ of the unmixed states as a function of $c$ . The other  mixing parameters have been fixed to $a=100$ MeV and $b=100$. The dotted line represents the unmixed glueball ($1750$), the dashed line represents the light unmixed meson ($1550$), the full line represents the heavy unmixed meson ($1900$)}
\label{threehMP}
\end{center}
\end{figure}
%%%%%%%%%%%%%%%%%%%%%%%%%%%%%%%%%%%%%%%%%%%%%%%%%%%%%%%%%

The first scenario requires very small mixing, as can be seen in Fig.\ref{threeMP} (left). The unmixed values  were chosen $m_{s1} = 1380$ MeV, $m_{s2} = 1530$ MeV and  $m_g = 1670$ MeV, the latter two standard deviations below  the central lattice value. The mixing parameters that fit the data are  small $a=b=50$ MeV and $c=70$ MeV.  Note that in this case  one can also  trivially construct a two state weak mixing scenario, with the outcome that  the $f_0(1710)$ is again mostly glueball, the $f_0(1500)$ mostly meson and the mixing parameters are small.Thus the $f_0(1370)$ is not required from the point of view of mixing.

I show in Fig.\ref{threeMP} (right) the probability distribution for the final mostly glueball state. It turns out that the experimental $f_0(1710)$ comes out mostly glueball in agreement with the lattice calculation of mixing \cite{Lee:1999kv,Vaccarino:1999ku} and other phenomenological analyisis \cite{Cheng:2009dg}.

%%%%%%%%%%%%%%%%%%%%%%%%%%%%%%%%%%%%%%%%%%%%%%%%%%%%%%%%%
\vskip 0.3cm
\begin{figure}[htb]
\begin{center}
\epsfig{file=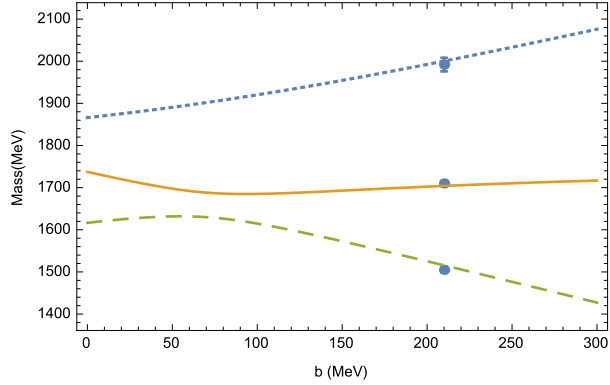,width=8.0cm,angle=0} 
\caption{Eigenvalues of the mixing matrix for the glueball and two meson states.  The masses and their errors are from PDG \cite{Agashe:2014kda}.}
\label{threehsM}
\end{center}
\end{figure}
%%%%%%%%%%%%%%%%%%%%%%%%%%%%%%%%%%%%%%%%%%%%%%%%%%%%%%%%%

Let me now elaborate on the other scenario, namely assuming $m_g > 1710$ MeV, i.e. I  take $1750$ MeV, close to the average value, as the unmixed glueball mass. As a result of Cardano's analysis, I know that the only way to get that mass down to the $f_0(1710)$ is by  incorporating a heavier meson. In particular, I show in Fig. \ref{threehMP} (left) a three state mixing where the $f_0(1710)$  is mostly glueball but the $f_0(2020)$, a not yet confirmed particle, must exist to get the experimental values for the lower masses. The corresponding probability curves,  Fig. \ref{threehMP} (right), show that the glueball component is at the level of 70\%. Most of the other 30 \% is carried by the $f_0(2020)$. Besides the existence of the $f_0(2020)$, another feature of this scheme is that the $f_0(1370)$, whatever it be, is not required, and thus the error bars are really very small and the fit quite restrictive. Another characteristic of this weak mixing scheme in the spread distribution of the unmixed spectrum: the particles appear with an energy step of $150-200$ MeV.  Note that in this case a two state mixing scenario could also  be trivially constructed, but here the $f_0(2020)$ would  be unavoidable and the $f_0(1500)$ would decouple from the mixing scheme.

\begin{figure}[htb]
\begin{center}
\epsfig{file=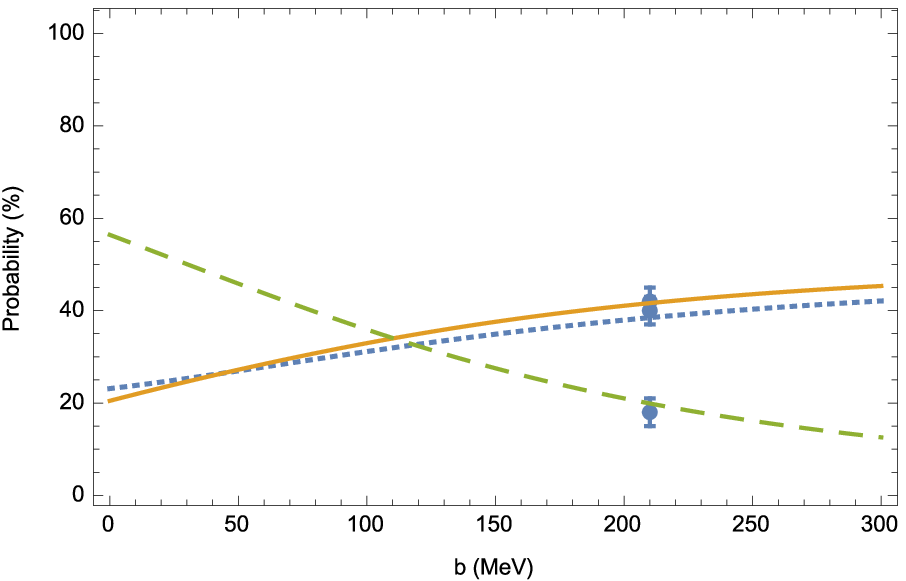,width=5.0cm,angle=0} \hspace{0.5cm}\epsfig{file=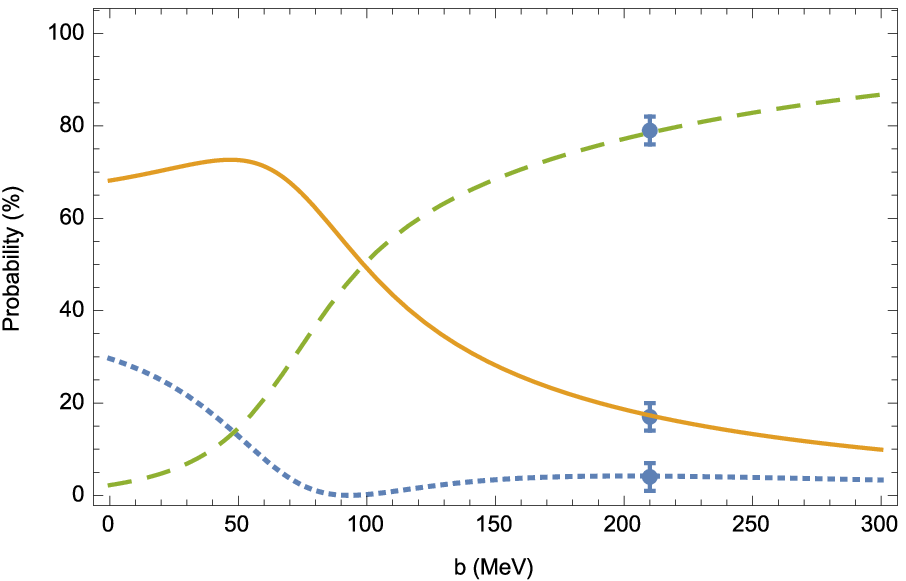,width=5.0cm,angle=0}  \hspace{0.5cm}\
\epsfig{file=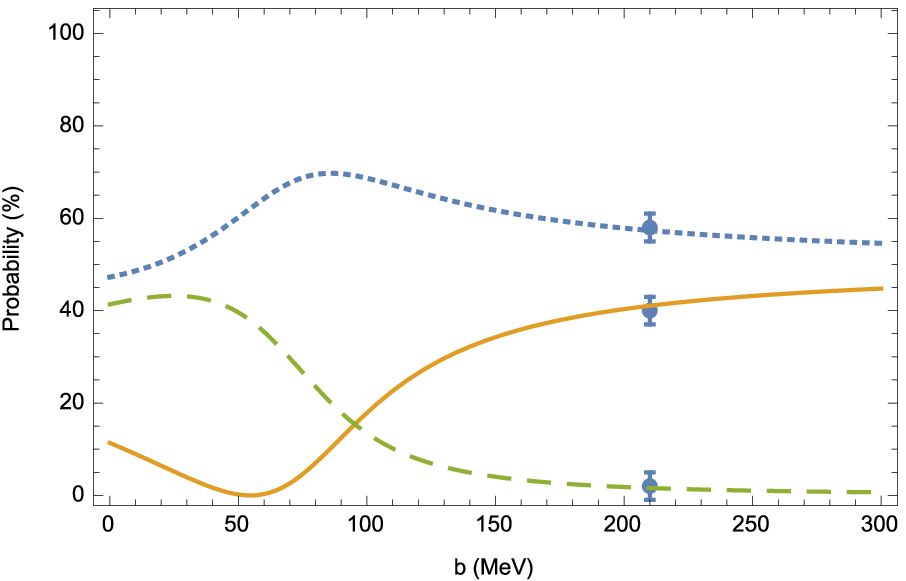,width=5.0cm,angle=0}  \\[0.5cm]
\caption{The probability distributions as a function of $b$ of the mixed states, for mixing parameters $a=100$ MeV, $c=70$ MeV. The notation follows the unmixed states: the full line represents the glueball ($1750$ MeV), the dashed line represents the meson ($1710$ MeV), the dotted line represents the meson ($1760$) MeV. The left figure represents the probability distribution of the $f_0(2020)$, the middle that of the $f_0(1710)$, and the right figure that of the $f_0(1500)$.}
\label{threehsP}
\end{center}

\end{figure}

Within the latter scenario also a strong mixing relization arises. As shown in Fig.\ref{threehsM}, I am  able to get a good fit to the data starting from almost degenerate unmixed states. In this case the unmixed values are for two mesons at $1710$ MeV and $1760$ MeV and a glueball at $1750$ MeV. The mixing parameters between the almost degenerate light meson and  glueball is large $b=210$ MeV, while the others are normal $a=100$ MeV, $c=70$ MeV. On the other hand this strong mixing affects the $f_0(1710)$ strongly, which now contains very little glueball, $\sim 17\%$, while increasing dramatically the glueball content of  the $f_0(1500)$, $\sim 41\%$, and  some that of the $f_0(2020)$, $\sim 42 \%$ , see Fig \ref{threehsP}. In this strong mixing scenario the glueball is a relic of the original glueball state and therefore it will be difficult to single out glueball properties. Again no need for the $f_0(1370)$ in the fit.

%%%%%%%%%%%%%%%%%%%%%%%%%%%%%%%%%%%%%%%%%%%%%%%%%%%%%%%%%

\vskip 0.3cm

%%%%%%%%%%%%%%%%%%%%%%%%%%%%%%%%%%%%%%%%%%%%%%%%%%%%%%%%%

I might summarize at this stage my findings by stating that accepting a mass value for a glueball as obtained from lattice QCD for the unmixed glueball state and allowing for mixing under the strict scrutiny of the $1/N_c$ expansion, two scenarios appear as dictated by the shape of the unmixed spectrum: i) a week mixing scenario, whose consequence is that the $f_0(1710)$ is an almost pure glueball state; ii) a strong mixing scenario in which the $f_0(1710)$ has almost no glueball component, while the $f_0(1500)$ and the $f_0(2020)$ have about $40 \%$ glueball component.   The role played by the scalar meson spectrum in the realization of mixing is fundamental. A detailed lattice QCD description of this spectrum would determine the mixing scenario. Within the weak mixing scenario it is very important the value of the mass of the unmixed glueball state. If it is below the $f_0(1710)$ the mixing proceeds via the $f_0(1500)$ and the $f_0(1370)$ and it is not very constrained due to the large indeterminacy in the latter. If that mass is above $1710$ MeV the $f_0(2020)$ is required to bring the mass value down, while the $f_0(1370)$ is unnecessary.

\section{Conclusions}
In this paper I take a very pragmatic point of view.
I assume that the lowest found lattice QCD glueball state is associated with real physical states
within the $f_0$ spectrum and try to establish the connection by implementing phenomenologically 
mixing with the nearby scalar mesons.  In order to estimate the mixings  I use the splitting pattern of the $f_0$ spectrum and $1/N_c$ arguments. 

I find  two possible weak mixing scenarios which 
associate the lattice glueball mostly with the $f_0(1710)$. In this case the exact value of the unmixed glueball state is fundamental to determine the physical realization of mixing, being the $f_0(1710)$ the dominant scale.  If the unmixed glueball mass is below that value, the $f_0(1370)$ should exist and may play a relevant role. If the mass is above, at least the $f_0(2020)$ is required to enter the mixing to bring down the unmixed mass to the physical mass. The scalar spectrum in both cases is loose, with an energy step between $150-200$ MeV.

I also find a strong mixing scenario which leads to the $f_0(1500)$ and the $f_0(2020)$ having a strong glueball component, while the $f_0(1710)$ is mostly  mesonic in character.  In this case the unmixed scalar spectrum presents an almost degeneracy around the unmixed glueball mass.

The main conclusion of this calculation is that given the fact that mixings are difficult to calculate in lattice QCD, a good knowledge of the scalar meson spectrum together with phenomenology will clarify the glueball constituency of the physical $f_0$'s in full QCD. Once the scalar spectrum is known the $f_0$ spectrum and the arguments about the mixing used in this communication will fix quite strongly the mixing parameters. 

Giving the above analysis I find  that the $f_0(1710)$ being mostly glueball is the most natural scenario. If it is accompanied, besides the $f_0(1500)$, by the $f_0(1370)$ or the $f_0(2020)$ in the mixing scenario is a matter of experimental determination.
It is clear that for this purpose not only masses, but also decays will be needed to determine the main properties of the glueball component. In here I have presented a guide of possible mixing schemes as characterized by the structure of  the spectrum.

\subsection*{Acknowledgments}

Correspondence with Jaume Carbonell and Vincent Mathieu  is
gratefully acknowledged. This work was supported  by the MINECO under contract FPA2013-47443-C2-1-P, by GVA-PROMETEOII/--2014/066, and by CPAN(CSD-00042).


\vskip 0.5cm



\begin{thebibliography}{99}

%\cite{Mathieu:2008me}
\bibitem{Mathieu:2008me}
  V.~Mathieu, N.~Kochelev and V.~Vento,
  %``The Physics of Glueballs,''
  Int.\ J.\ Mod.\ Phys.\ E {\bf 18} (2009) 1
  [arXiv:0810.4453 [hep-ph]].
  %%CITATION = ARXIV:0810.4453;%%

%\cite{Lee:1999kv}
\bibitem{Lee:1999kv}
  W.~J.~Lee and D.~Weingarten,
  %``Scalar quarkonium masses and mixing with the lightest scalar glueball,''
  Phys.\ Rev.\ D {\bf 61} (2000) 014015
  [hep-lat/9910008].
  %%CITATION = HEP-LAT/9910008;%%

%\cite{Vaccarino:1999ku}
\bibitem{Vaccarino:1999ku}
  A.~Vaccarino and D.~Weingarten,
  %``Glueball mass predictions of the valence approximation to lattice QCD,''
  Phys.\ Rev.\ D {\bf 60} (1999) 114501
  [hep-lat/9910007].
  %%CITATION = HEP-LAT/9910007;%%

%\cite{Morningstar:1999rf}
\bibitem{Morningstar:1999rf}
  C.~J.~Morningstar and M.~J.~Peardon,
  %``The Glueball spectrum from an anisotropic lattice study,''
  Phys.\ Rev.\ D {\bf 60} (1999) 034509
  [hep-lat/9901004].
  %%CITATION = HEP-LAT/9901004;%%

%\cite{Chen:2005mg}
\bibitem{Chen:2005mg}
  Y.~Chen, A.~Alexandru, S.~J.~Dong, T.~Draper, I.~Horvath, F.~X.~Lee, K.~F.~Liu and N.~Mathur {\it et al.},
  %``Glueball spectrum and matrix elements on anisotropic lattices,''
  Phys.\ Rev.\ D {\bf 73} (2006) 014516
  [hep-lat/0510074].
  %%CITATION = HEP-LAT/0510074;%%


%\cite{Gregory:2012hu}
\bibitem{Gregory:2012hu}
  E.~Gregory, A.~Irving, B.~Lucini, C.~McNeile, A.~Rago, C.~Richards and E.~Rinaldi,
  %``Towards the glueball spectrum from unquenched lattice QCD,''
  JHEP {\bf 1210} (2012) 170
  [arXiv:1208.1858 [hep-lat]].
  %%CITATION = ARXIV:1208.1858;%%



%\cite{Amsler:1995td}
\bibitem{Amsler:1995td}
  C.~Amsler and F.~E.~Close,
  %``Is f0 (1500) a scalar glueball?,''
  Phys.\ Rev.\ D {\bf 53} (1996) 295
  [hep-ph/9507326].
  %%CITATION = HEP-PH/9507326;%%


%\cite{Barberis:1996iq}
\bibitem{Barberis:1996iq}
  D.~Barberis {\it et al.}  [WA102 Collaboration],
  %``A Kinematical selection of glueball candidates in central production,''
  Phys.\ Lett.\ B {\bf 397} (1997) 339.
  %%CITATION = PHLTA,B397,339;%%

%\cite{Cheng:2009dg}
\bibitem{Cheng:2009dg}
  H.~Y.~Cheng,
  %``Scalar and Pseudoscalar Glueballs,''
  Int.\ J.\ Mod.\ Phys.\ A {\bf 24} (2009) 3392
  [arXiv:0901.0741 [hep-ph]].
  %%CITATION = ARXIV:0901.0741;%%
  
 %\cite{Agashe:2014kda}
\bibitem{Agashe:2014kda}
  K.~A.~Olive {\it et al.}  [Particle Data Group Collaboration],
  %``Review of Particle Physics,''
  Chin.\ Phys.\ C {\bf 38} (2014) 090001.
  %%CITATION = CHPHD,C38,090001;%%
  
\end{thebibliography}
\end{document}